# Optimization of Sequential Microwave-Ultrasound-Assisted Extraction for Maximum Recovery of Quercetin and Total Flavonoids from Red Onion (*Allium cepa* L.) Skin Wastes

Zeinab Jabbari Velisdeh[1], Ghasem D. Najafpour[1*], Maedeh Mohammadi[1], Fatemeh Poureini[1]

[1] Biotechnology Research Laboratory, Department of Chemical Engineering, Babol Noshirvani University of Technology, Babol, Iran

[*] **Corresponding author:** Prof. G.D. Najafpour, Biotechnology Research Laboratory, Faculty of Chemical Engineering, Babol Noshirvani University of Technology, Babol, Iran. E-mail: najafpour@nit.ac.ir

**Acknowledgments**

We gratefully acknowledge Biotechnology Research Laboratory, Babol Noshirvani University of Technology for facilitating the research and providing the necessary support to conduct the present work.

**Highlights**

- Extraction of high quality quercetin from red onion skin using MUAE method
- MUAE recovered significantly higher total flavonoids than other methods
- Optimization of the extraction factors of MUAE using Box-Behnken design
- Ultrasound temperature is a highly significant extraction factor
- MAE is a desired choice as a pre-treatment before UAE for the recovery process

**Abstract**

This study aimed to optimize the extraction conditions in order to maximize the recovery yields of quercetin and total flavonoids from red onion skin wastes using sequential microwave-ultrasound-assisted extraction. Five effective factors on quercetin extraction yield were investigated using response surface methodology. The method was successfully performed under optimal conditions of 60 s microwave irradiation followed by 15 min sonication at 70 °C, 70% ethanol with solvent to solid ratio of 30 mL/g. Based on the optimization results, ultrasound temperature was found to be a highly significant and influential factor for the recovery. The maximum recovery yields of quercetin and total flavonoids from red onion skin were estimated to be 10.32 and 12.52%, respectively. The predicted values for quercetin (10.05%) and total flavonoids (12.72%) were very close to the experimental results. The recovery yields obtained from different extraction methods were as follows: ultrasound-microwave-assisted extraction (7.66% quercetin and 10.18% total flavonoids), ultrasound-assisted extraction (5.36% quercetin and 8.34% total flavonoids), and microwave-assisted extraction (5.03% quercetin and 7.91 % total flavonoids). The validity of the projected model was examined by the obtained experimental data; in which, the validated model was suitable for the recovery of the valuable products from onion skin wastes for further scale-up in the food processes.

1. **Introduction**

Quercetin, with the molecular formula $C_{15}H_{10}O_7$ and molecular weight of 302.236 g/mol (Fig. 1), is known as a natural flavonoid (polyphenolic group) presenting in various parts of onions (*Allium cepa* L.). In addition to the anti-inflammatory and antibacterial properties of polyphenols such as quercetin, it has been proposed to have beneficial effects on cancer treatment, diabetes and obesity [1- 4]. Onion skin as an industrial waste is produced from different food processing plants and culinary services [5]. Although being a waste, onion skin contains significant amounts of valuable compounds, including natural flavonoids such as quercetin and kaempferol [6]. Among different varieties of onion, red and yellow ones have the major flavonoid contents [7]. Studies have shown that the outer layer (skin) of onion has a higher quercetin content than the inner and middle layers [8].

**Fig. 1**

Conventional solvent extraction methods suffer from deficiencies such as high amount of solvent consumption, long extraction time and thermal degradation of some sensitive compounds, hence modern extraction methods have been utilized recently for the recovery of various compounds from plant materials [9, 10]. For the recovery of the flavonoid compounds like quercetin from onion skin, various extraction techniques such as subcritical fluid extraction (SFE) [11], conventional solvent extraction [12], microwave-assisted extraction (MAE) and ultrasound-assisted extraction (UAE) [13] have been employed. However, the reported extraction yield of flavonoids using SFE and MAE are quite low [11, 13]. Among the advance extraction methods, UAE has always proved to be an efficient and eco-friendly method due to its short extraction time and low solvent consumption for the recovery of flavonoids [14, 15]. Ultrasonic waves disrupt the cell wall by creating cavitation bubbles, which then grow and explode, thus facilitate the process of diffusion of the solvent into the plant cell and extraction of the targeted compound [16, 17]. However, the high ultrasonic temperature and long ultrasonication time can destroy the structure of sensitive compounds [18].

Therefore, use of an advanced method for the rapid and effective extraction of flavonoids from onion skin wastes is needed. While the combination of MAE and UAE as a sequential technique is one of the most efficient and speedy extraction methods [19], to the best of our knowledge, no previous studies have been conducted on the use of sequential microwave-ultrasound-assisted extraction (MUAE) for the recovery of quercetin from onion skin. To date, several research studies have reported the extraction of valuable products from plant materials using UAE and MAE as a sequential technique such as sage by-products for antioxidants recovery [20], natural colorants from sorghum husk [19], Piperine from Black Pepper [21] and Galactomannan from fenugreek seed [22].

Microwave-assisted extraction as an eco-friendly, effective and green extraction technique, uses electromagnetic energy and direct heating to evaporate volatile compounds in the plant sample, which causes cell deformation and allows the solvent to penetrate into the cell wall, thus preparing the cell for the subsequent extraction process [23-26]. Exposure of these inflamed cells immediately after MAE to UAE can completely destroy the cell wall and thus extract the desired compounds with higher yield and efficiency. We hypothesized that combining MAE and UAE could shorten the extraction time and reduce the possibility of the decomposition of flavonoids structure, hence enhance both the stability and recovery yield of the extracted compounds (Fig. 2). Therefore, MAE can be considered as an effective pre-treatment step before the main treatment (UAE) for the recovery process [27]. However, the prolonged exposure to microwave irradiation and ultrasonic waves can destroy flavonoids; but, an optimized exposure time can dramatically increase the extraction yield.

**Fig. 2**

The aim of the present study was to enhance the recovery yields of quercetin and total flavonoid from red onion skin by optimizing the extraction condition and investigating the influential extraction factors. For this, the effects of different factors of each method, such as solvent type, solvent concentration, solvent-to-solid ratio, microwave power, microwave irradiation time, ultrasound power, ultrasound time, ultrasound temperature, ultrasound frequency and particle size on the extraction yield were investigated. Various experiments were performed for two separate sets of experiment: single-factor design and the response surface methodology (RSM). RSM was used as a statistical and mathematical tool that explains the interaction between significant factors and their effects on the extraction process [28, 29]. Furthermore, the Box-Behnken Design (BBD) was applied to define the extraction conditions. The recovery yield of quercetin and total flavonoids extracted by MUAE, UMAE, UAE and MAE were compared to specify the efficacy of MUAE.

2. **Materials and methods**

## 2.1. Raw materials and solvents

Red onion skins were obtained from a local grocery market (Amol, Iran). The collected skins were washed with cold distilled water and subsequently dried at 130 °C for 15 min. The ash and moisture contents of the skins (g/100 g) were 8.7 ± 0.84 and 8.9 ± 0.42%, respectively. The dried onion skins (DOS) were ground using an electric grinder and sieved through mesh no. 50 to 140 (100-600 μm) using standard sieves, then stored in glass jars and kept in a cold room (–4 ºC) for further use. Chemicals such as potassium acetate (> 99%), aluminum chloride (> 95%), quercetin (> 95%), isorhamnetin (> 95%) and kaempferol (> 97%) standards were purchased from Sigma-Aldrich (St. Louis, Mo, USA). Solvents such as ethanol, acetone, water and ethyl acetate (> 99.9%) were obtained from Scharlau (Barcelona, Spain). HPLC grade methanol, orthophosphoric acid and water were purchased from Merck (Darmstadt, Germany).

## 2.2. Sequential microwave-ultrasound-assisted extraction

For the extraction process, a commercial microwave oven (Samsung, CQ4250, Korea) with a frequency of 2.45 GHz and maximum input power of 1150 W was used. One gram of DOS was thoroughly mixed with a specific amount of the desired solvent in an Erlenmeyer flask. To prevent any possible thermal degradation of the sample and also the loss of solvent in the course of MAE, an intermittent heating-cooling process was used where after 10 s of irradiation, the flask was placed in an ice-bath and cooled down. This intermittent heating-cooling process was continued until the specified extraction time was passed. A digital temperature controller (Autonics, TCN4L-24R, South Korea) was used to specify the actual temperature of the sample during the experiment. Then, the flask was closed with a rubber stopper and immersed in an ultrasonic bath (Elmasonic, P30H, Germany) and sonicated. For efficient sonication, the level of water in the bath was always above the level of solvent in the immersed flask. To prevent the excessive heating of the sample during the UAE process, the sample was cooled by adding ice cubes into the water bath, and the temperature was continuously monitored during the process. The manufacturer rated the apparatus with a peak power and effective power rating of 480 and 120 W, respectively; it had a proprietary algorithm for power adjusting based on system impedance. At a fixed bath volume (1.4 L) for 50, 60, 70, 80, 90, and 100% power settings, the peak powers were 171, 206, 240, 274, 309, and 343 $W/cm^2$, respectively, and the ultrasonic effective powers were 42, 51, 60, 69, 77, and 86 $W/cm^2$, respectively. After sonication, the sample was filtered through a Whatman filter paper (0.45 μm). The filtrate was dried in an oven at 60 ºC for 12 h till all the solvent evaporated. The dried sample was dissolved in 1 mL of methanol and placed in a microtube and centrifuged by a microcentrifuge (Labnet, C1301-P, Korea) at 6000 rpm for 5 min. The samples were then analyzed to identify quercetin and total flavonoids concentration using a high-performance liquid chromatograph (HPLC). The optimal points at which the highest extraction yield was obtained were determined for the sequential extraction method (MUAE). The results were compared with those obtained by ultrasound-microwave-assisted extraction (UMAE) and the single extraction methods (MAE and UAE). The amounts of quercetin and total flavonoids contents (TFC) were calculated based on a standard calibration curve developed using different concentrations of quercetin (1 to 100 mg/L) in methanol.

The extraction yields were calculated by Eq. (1):

$$\text{Yield (\%)} = \frac{\text{Weight of extracted quercetin or total flavenoid (mg)}}{\text{Weight of DOS (g)}} \times 100 \quad (1)$$

### 2.3. Total flavonoids content (TFC)

The total flavonoid content of onion skin extract was determined according to the colorimetric method [30]. For this, 0.5 mL of extract solution was mixed with 1.5 mL of methanol, followed by the addition of 0.1 mL (10%) aluminum chloride, 0.1 mL potassium acetate (1 M), and 2.8 mL distilled water. Then, to the mixture was incubated at room temperature for 30 min. The absorbance of the sample was recorded at 370 nm using a spectrophotometer (Analytik Jena AG, SPEKOL 1500, Germany). TFC was determined according to the standard calibration curve of quercetin and expressed as quercetin equivalents (mg QE/g DOS). Furthermore, for the analysis of quercetin, the absorbance of the extracted sample was read at a wavelength of 370 nm. All experiments were performed in triplicates.

### 2.4. High performance liquid chromatography (HPLC)

HPLC (Smartline, Knauer, Germany) equipped with a Eurospher I 100-5 C18 column with dimensions of 250 × 4.6 mm and UV detector 2500 series was used. The column temperature was 30 °C. The detector was set to scan from 200 to 600 nm for monitoring retention time of flavonoids, and two wavelengths for the detection of target compounds were at 295 and 365 nm. The injection volume was 20 μL and the mobile phases were: 0.5% of orthophosphoric acid in water (solvent A) and methanol (solvent B) with a solvent gradient elution program: 40 to 60% solvent B from 0 to 10 min, 60% B from 10 to 21 min, 60 to 40% B from 21 to 23 min, 40% B from 23 to 26 min, then was held at 40% solvent B until the end of the run at 30 min for column equilibration [31]. The elution flow rate was 1.0 mL/min.

### 2.5. Single-factor design

Single-factor design was applied to study the effects of different factors (solvent type,

ethanol concentration, solvent-to-solid ratio, microwave power, microwave irradiation time, ultrasound power, ultrasound time, ultrasound temperature, ultrasound frequency and particle size) on the MUAE of quercetin from DOS. Different extraction solvents including ethanol, acetone, water and ethyl acetate were tested for MUAE. In addition, the effects of ethanol as the best performing solvent at volume fractions of 50, 60, 70, 80, 90, and 100% (in water as a co-solvent) and several solvent-to-solid ratios (10:1, 20:1, 30:1, 40:1, 50:1 and 60:1 mL/g) were investigated. For MAE, microwave powers of 100, 180, 300, and 450 W and microwave irradiation times of 30, 60, 90, 120, 150, and 180 s were investigated. For UAE, the ultrasound powers of 50, 60, 70, 80, 90 and 100% of full power, sonication times of 10, 15, 20, 25, 30, 35, and 40 min, ultrasound temperatures of 30, 40, 50, 60, 70 and 80 °C and ultrasound frequencies of 37 and 80 kHz were experimented. Besides, for the case of MUAE, different particle sizes from 100 to 600 μm were used. In the course of experiment, when one independent factor was changed, the other experimental factors were kept constant. In addition, the defined optimal points based on the results of single-factor experiments were selected as the central points of the RSM experiments. For comprehensive examination, the lower and upper levels of the optimal points were applied in RSM runs.

*2.6. Experimental design for RSM*

In order to optimize the MUAE process, RSM as an optimization method was employed to determine the optimum conditions for the maximum recovery yield of quercetin and total flavonoids from red onion skin wastes using Box-Behnken Design (BBD). The main factors for RSM design were selected based on the most effective factors on the results obtained from the single-factor experiments. The five effective variables on the extraction including ultrasound time (min, $X_1$), ultrasound temperature (°C, $X_2$), solvent-to-solid ratio (mL/g, $X_3$) microwave time (s, $X_4$), and ethanol concentration (%, $X_5$) are summarized in Table 1 at three levels (-1, 0, +1), with $X_1$ (15, 30, and 45 min), $X_2$ (60, 70, and 80 °C), $X_3$ (20, 30, and 40 mL/g), and $X_4$ (30, 60, and 90 s) and $X_5$ (50, 60, and 70%). In general, 46 experimental runs (in triplicates) were carried out. The dependent variables (responses) were quercetin ($Y_1$) and total flavonoids ($Y_2$) yields. The design condition and results of the 46 experimental runs to optimize quercetin and total flavonoids extraction conditions are summarized in Table 2. According to the variance analysis of the regression model, a p-value less than 0.05 indicates the significance of terms and a p-value less than 0.0001 indicates the high significance of the regression model terms. Moreover, p-values greater than 0.10 indicate that the model terms are insignificant. The regression analysis of the extraction responses was fitted using a second-order polynomial equation as expressed in Eq. (2):

$$Y_n = \beta_0 + \sum_{i=1}^{k} \beta_i x_i + \sum_{i=1}^{k} \beta_{ii} x_i^2 + \sum_{i}^{k-1} \sum_{j}^{k} \beta_{ij} x_i x_j \qquad (2)$$

where $Y_n$ represents the extraction responses, $β_0$ is the intercept, $β_i$ denotes the coefficient of the linear, $β_{ii}$ is the quadratic term and $β_{ij}$ represents the cross-product term, $x_i$ and $x_j$ denote the independent variables and $k$ is the number of variables ($k=5$).

**Table 1**

*2.7. Statistical analysis*

The statistical analyses of data obtained from one-factor experiments were performed by one-way analysis of variance (ANOVA) and Tukey's test. Each experiment was carried out in triplicate and the results were expressed as the mean values ± SD. Optimization and modeling of RSM for the recovery process were performed using the software Design-Expert (Version 7.0.0, Stat-Ease Inc., USA). The significance of independent factors was specified using Analysis of Variance (ANOVA).

## 3. Results and discussion

*3.1. Single-factor analysis of quercetin extraction*

The effect of the type of solvent on the extraction yield of quercetin in MUAE was investigated using four solvents, including ethanol, acetone, water and ethyl acetate. Other parameters including microwave irradiation time (120 s), ultrasonication time (15 min), ultrasound temperature (60 °C), solvent-to-solid ratio (50 mL/g), microwave power (180 W), ultrasound power (100%), ultrasound frequency (80 kHz) and particle size (420 µm) were set constant. The results in Fig. 3A show that the four solvents had different effects on the extraction yield under the same extraction conditions due to different solvent properties. The maximum extraction yield was obtained using ethanol (3.50%), followed by acetone (2.00%), water (1.15%) and ethyl acetate (0.96%). A solvent with very low vapor pressure is desired in UAE because a low vapor pressure solvent creates few cavitation bubbles but with a stronger collapse force, which leads to enhanced destruction of the plant cell wall [32]. Ethanol has the lowest vapor pressure (59.02 mmHg) among the used solvents at room temperature, and this explains the better extraction yield obtained by this solvent compared to the other two investigated solvents (acetone (229.52 mmHg) and ethyl acetate (73.00 mmHg)). Even though water is a proper solvent for the flavonoids extraction and has a lower vapor pressure (23.80 mmHg) in comparison with ethanol [33]. but in the case of MAE, ethanol has a higher dissipation factor (tanδ = 0.94) compared to water (tanδ = 0.12) [21]. Tanδ shows the ability of the solvent to convert microwave energy into thermal energy [34]; therefore, it was expected that a higher yield could be obtained using ethanol due to its high ability to provide more heat into the extraction medium. Accordingly, ethanol with a maximum yield of 3.50 ± 0.40%, was chosen as the best extraction solvent for the recovery of bioactive compounds from DOS.

The effect of the concentration of ethanol mixed with water on the extraction yield of quercetin at the volume fractions of 50, 60, 70, 80, 90 and 100% was investigated. As shown in Fig. 3B, when the mixing ratio was increased from 50 to 60%, a significant increase in the extraction yield was observed; this was followed by a decrease with an increase of the mixing ratio to above 60%. The average temperature changes are also shown in Fig. 3B. The addition of water to ethanol can cause the swelling of solid particles in the solvent medium and provide further contact area between solvent and solid. Therefore, it would result in a better mass-transfer and dissolution of the targeted component [35]. Thus, the ethanol concentration of 60% with the maximum yield of 4.70 ± 0.11% was selected as the optimum ethanol concentration for subsequent experiments. For extensive investigation on the ethanol concentration, the range of 50-70% was chosen for RSM runs.

The effect of microwave irradiation time on the extraction yield is illustrated in Fig. 3C. As the extraction results show, a significant rise in the extraction yield was observed with prolonging the irradiation time from 30 to 60 s and after that, a reducing trend from 60 to 180 s was observed. During the extraction process, despite the risk of flavonoid degradation at longer microwave irradiation time, MAE can easily disrupt the plant's cell wall and release the desired components [36]. Therefore, 60 s with a peak yield of 5.01 ± 0.02% was selected as the optimum extraction time for subsequent experiments. For further examination, the microwave irradiation time range of 30-90 s was selected to be applied in the RSM runs.

Fig. 3D depicts the effect of microwave power on the extraction yield. The extraction yield increased from 4.97 to 5.01%, with increasing the microwave input power from 100 to 180 W. However, the extraction yield significantly decreased with raising the input power to beyond 180 W. The average temperature changes with variations of the microwave power during the experiment are also shown in Fig. 3D. The rise of the microwave power from 100 to 450 W induced an increase in the average temperature from around 29 to 86 ºC. A possible explanation for the observed trend in the extraction yield is that at high microwave powers, thermal degradation of phytochemical structures might occur and hence reduce the extraction yield [37]. Hence, the highest amount was observed at an input power of 180 W with a maximum yield of 5.01 ± 0.04%, which was chosen as the optimum microwave power for quercetin extraction from DOS.

The effect of sonication time on the extraction yield of quercetin from DOS was investigated,

and the results are depicted in Fig. 3E. During the first 30 min of the ultrasonication, the extraction yield significantly increased to 5.86%, but with an increase of the sonication time to 40 min no significant changes in the extraction yield were observed. At the beginning of the ultrasound extraction process, ultrasonic waves can create a high level of energy due to the collapse of the cavitation bubbles and rapid disruption of the cell wall of the plant sample, which enhances the rate of extraction due to the large contact area between the plant sample and the extraction solvent [38]. However, as the sonication time increases with the continuous release of the targeted components, the extraction solvent becomes saturated and thus the driving force for the mass transfer reduces. Therefore, considering the commercial aspects and energy utilization, 30 min with a yield of 5.86 ± 0.02% was selected as the optimum ultrasonication time for subsequent experiments. Accordingly, for a comprehensive consideration, the ultrasound time range of 15-45 min was selected for further use in the RSM runs.

The effect of ultrasound frequency on the extraction yield was specified for two ultrasound frequencies of 37 and 80 kHz. As results in Fig. 3F show, a significant decrease in the extraction yield was experienced with the change of the frequency from 37 to 80 kHz. The probable reason is that at high frequencies, there is no opportunity for cavitation bubbles to grow [36]. In addition, when using an 80 kHz frequency, a longer sonication time is required. Thus, the frequency of 37 kHz with a yield of 7.52 ± 0.03% was selected as the optimum ultrasound frequency for the extraction of quercetin from DOS.

Fig. 3G illustrates the effect of ultrasonic power on the extraction yield of quercetin. The ultrasound power had a positive effect on the quercetin yield where raising the power from 50 to 100% led to an increase in the extraction yield from 6.46 to 7.52%. The highest yield was obtained at the highest ultrasound power (100%). Increase of the ultrasound power can generate more energy and better destroy the cell structure due to the collapse of cavitation bubbles, which can release bioactive compounds [39]. Therefore, the ultrasonic power of 100% with the extraction yield of 7.52 ± 0.08% was chosen as the optimum power for the extraction of quercetin from DOS.

The results exhibited in Fig. 3H show the effect of ultrasound temperature on the extraction yield of quercetin. An increase in the ultrasound temperature from 30 to 70 ºC resulted in an increase in the extraction yield from 6.44 to 8.01%; however, further increase of the temperature to 80 ºC reduced the extraction yield of quercetin to 6.93%. Despite the fact that low ultrasonic temperature can retain thermo-sensitive compounds, the thermal effect is weak and the mass transfer of the plant particles into the liquid phase is not high, but at high ultrasound temperatures, cell disruption and dissolution increase which enhance the extraction yield [40]. In the current study, at temperatures above 70 ºC, the extraction yield reduces. An explanation for this phenomenon is that the boiling point of the extraction solvent (ethanol) is 78 ºC, and the sonochemical effects due to cavitation bubbles collapse decrease when the extraction temperature is close to the solvent boiling point [36]. Also, extreme extraction temperature limits biological function and may degrade the structure of the targeted compounds during the experiment [41]. So, the temperature of 70 ºC with the yield of 8.01 ± 0.10% was selected as the optimum temperature for further experiments. Therefore, for broader assessment, the ultrasound temperature range of 60-80 ºC was chosen for subsequent use in the RSM runs.

The results in Fig. 3I indicate the effect of particle size on the extraction yield of quercetin. As observed, particle size had an insignificant effect on the extraction yield and the yield changed only slightly with variations in the particle size of DOS. The extraction yield reached its maximum value at the size of 180 μm and reduced slightly when the particle size was decreased to 150 μm. However, from 150 to 100 μm, the yield remained almost constant. Also, for particles above 250 μm, the yield remained almost unchanged. The constant trend observed for small particles can be due to two counteracting effects. Despite the fact that small particles have a large surface area that can increase the extraction yield, these particles tend to stay on the solvent surface, so the ultrasonic waves are limited during the extraction process, which reduce the extraction yield. On the other hand, in the case of too large particles, the diffusion of the extraction solvent and release of the targeted component to the liquid medium are lower due to the smaller available surface, reducing the extraction yield [42]. Accordingly, 180 μm with a yield of 8.23 ± 0.02% was chosen as the best particle size for the extraction of quercetin from DOS.

The effects of solvent-to-solid ratio on the extraction yield with the average temperature changes are shown in Fig. 3J. The results indicated an increasing trend for an increase in the solvent-to-solid ratio from 10 to 30 mL/g. After that, the quercetin yield began to decrease with an increase of the ratio to 60 mL/g. The explanation for this behavior is that the use of large amounts of solvent may raise the extraction of undesirable compounds and accordingly decrease the purity of the desired compound. In contrast, at low solvent-to-solid ratios, the required amount of solvent for diffusion into the plant structure to extract the targeted compounds may not be available. Hence, in order to obtain the best extraction yield in the extraction process, a balance must be established between the amount of solvent and the solid. Therefore, it is highly important to reduce solvent consumption and increase the recovery yield simultaneously [43]. In general, the greater amount of solvent has a positive effect on extraction yield due to more contact between the sample and the solvent, as it can dissolve and extract more plant components [44], but also can extract undesired components that can reduce the purity and extraction yield of the targeted compound. There was no significant difference between the extraction yield of 30:1 mL/g (8.45%) and 40:1 mL/g (8.30%). With consideration of the economic aspects and investigation of the results, the ratio of 30:1 mL/g with a yield of 8.45 ± 0.05% was selected as the optimum ratio for further experiments. For a comprehensive evaluation, the range of solvent-to-solid ratio was selected from 20:1 to 40:1 mL/g for use in the RSM runs.

**Fig. 3**

**Table 2**

3.2.1. *Effect of MUAE parameters on the recovery yield of quercetin from DOS*

The analysis of variance (ANOVA) of response 1 ($Y_1$) is summed up in Table 3. The model p-value of < 0.0001 indicates the high significance of the regression model. The lack of fit result was insignificant with the p-value of 0.0515, which implies a significant fitting of the regression model with the data. The model terms of $X_1X_2$, $X_2^2$, $X_3^2$ and $X_5^2$ were highly significant ($p < 0.0001$). The terms $X_1X_4$, $X_1X_5$, $X_2X_4$, $X_2X_5$, $X_1^2$, $X_4^2$ were significant ($p < 0.05$). The other terms were insignificant ($p > 0.10$). The variance analysis of the model indicated the ultrasound temperature, which had three terms with significant interactions ($X_1X_2$, $X_2X_4$ and $X_2X_5$) and ultrasound time with three significant interaction terms ($X_1X_2$, $X_1X_4$ and $X_1X_5$) followed by ethanol concentration with two significant interaction terms ($X_1X_5$ and $X_2X_5$) were the most highly significant terms for recovery of quercetin from DOS. The quadratic equation relating time and temperature of ultrasound, solvent-to-solid ratio, microwave time and ethanol concentration is shown in Eq. (3):

$$Y_1 = 8.41 - 0.99X_1 - 0.65X_2 + 0.42X_5 + 0.39X_1X_2 + 0.16X_1X_4 - 0.17X_1X_5 - 0.25X_2X_4 - 0.24X_2X_5 + 0.11X_1^2 - 0.79X_2^2 + 0.32X_3^2 + 0.14X_4^2 + 0.31X_5^2 \quad (3)$$

where $Y_1$ is the yield of quercetin, $X_1$, $X_2$, $X_3$, $X_4$ and $X_5$ are the coded variables for ultrasound time, ultrasound temperature, solvent-to-solid ratio, microwave time and ethanol concentration, respectively.

The correlation coefficient $R_1$ for the statistical relationship between actual and predicted points was measured. The $R_1$ for response 1 with an absolute value of 98.55% indicates a very strong relationship and, the adjusted $R_1$ of 97.40% for $Y_1$ indicates a high influence of experimental factors. Also, for $Y_1$, the predicted $R^2$ of 94.41% is in reasonable agreement with the adjusted $R^2$.

The quercetin yield for all the 46 extraction tests was in the range of 6.51 –10.32%. The maximum extraction yield for quercetin (10.32%) was achieved at the conditions of run number 28, with ultrasound time of 15 min, ultrasound temperature of 70 ºC, solvent-to-solid ratio of 30 mL/g, microwave time of 60 s, and ethanol concentration of 70%. In addition, the minimum extraction yield of quercetin from the MUAE method (6.51%) was obtained from run number 17.

### 3.2.2. Effect of MUAE parameters on the recovery yield of total flavonoids from DOS

The ANOVA of response 2 ($Y_2$) is summed up in Table 3. The model p-value of <0.0001 implies the high significance of the regression model. An insignificant result in the lack of fit with a P-value of 0.1276 indicated the model fitted the data very well. In this case, the model terms of $X_1$, $X_2$, $X_5$, $X_1X_2$, $X_2^2$, $X_3^2$ and $X_5^2$ were highly significant ($p < 0.0001$). The terms of $X_1X_4$, $X_2X_4$, $X_2X_5$, $X_1^2$ were significant ($p < 0.05$). The other terms were insignificant ($p > 0.10$). For recovery of total flavonoids from DOS, the variance analysis of the model based on Table 2 suggested ultrasound temperature with three important interaction terms ($X_1X_2$, $X_2X_4$ and $X_2X_5$) and the ultrasound time with two terms of interactions ($X_1X_2$, $X_1X_4$) followed by ethanol concentration with one interaction term ($X_2X_5$) were the most highly significant terms for recovery of total flavonoids from DOS. The quadratic equation relating time and temperature of ultrasound, solvent-to-solid ratio, microwave time and ethanol concentration is shown in Eq. (4):

$$Y_2 = 10.42 - 1.02X_1 - 0.68X_2 + 0.36X_5 + 0.64X_1X_2 + 0.19X_1X_4 - 0.23X_2X_4 - 0.26X_2X_5 + 0.17X_1^2 - 0.58X_2^2 + 0.27X_3^2 + 0.30X_5^2 \quad (4)$$

where $Y_2$ is the yield of total flavonoids, $X_1$, $X_2$, $X_3$, $X_4$ and $X_5$ are the coded variables for ultrasound time, ultrasound temperature, solvent-to-solid ratio, microwave time and ethanol concentration, respectively.

For the second response ($Y_2$), the correlation coefficient $R^2$ and adjusted $R^2$ of the model were evaluated. $R^2$ with a value of 0.9812 indicates a good prediction for the response model and the adjusted $R^2$ with a value of 0.9661 also represents a significant adjustment for the model. The $R^2$ and adjusted $R^2$ indicate a close relationship between the predicted and actual values. Also, in this case for $Y_2$, the predicted $R^2$ of 0.9287 is in reasonable agreement with the adjusted $R^2$.

The total flavonoids yield for all the 46 extraction tests was in the range of 8.84–12.52%. The highest extraction yield for total flavonoids (12.52%) was achieved at the conditions of run number 1, with ultrasound time of 15 min, ultrasound temperature of 60 ºC, solvent-to-solid ratio of 30 mL/g, microwave time of 60 s, and ethanol concentration of 60%. In addition, the minimum yield of total flavonoids from MUAE (8.84%) was achieved in run number 17.

**Table 3**

### 3.2. Interactions between MUAE factors of response surface

The three-dimensional response surfaces and contour plots for quercetin extraction demonstrate the effects and interactions between the extraction factors (Fig. 4A–D). The surface showed an interaction between ultrasound temperature and ultrasound time and indicated that the extraction yield of quercetin decreased with an increase in ultrasound temperature and time at the same time (Fig. 4A). It is speculated that by increasing the time of ultrasonication and the continuous release of the product, the extraction solvent will be saturated with the extracted products [45]. In addition, an increase of the temperature can accelerate the process of solvent evaporation. It can reduce the contact surface between the solvent and the solid, thereby decreasing the extraction yield of quercetin.

Fig. 4B examines the interactions between the two parameters of ethanol concentration and ultrasound time. Considering the ethanol concentration at an ultrasound time of 5 min, increasing the ethanol concentration increased the yield of quercetin extraction. However, in general, the simultaneous decrease in both parameters indicated an increase in the maximum value of extraction yield. The possible reason is that by reducing the rate of evaporation of the solvent due to decreasing temperature, and also by reducing the content of the desired solvent and adding water as a co-solvent, the swelling degree of the onion skins in the solvent medium may increase and enhance the surface of contact between the solvent and the solid [46]. Hence, it can raise the surface of yield.

The response surface in Fig. 4C shows an effective interaction between microwave time and ultrasound temperature. Nevertheless, microwave time alone did not have a significant effect on the quercetin extraction yield. At the ultrasound temperature of 60 ºC, as the microwave time decreased, the extraction yield also decreased. However, at an ultrasound temperature of 80 ºC, reducing the microwave time increased the surface of the extraction yield. In general, increase of these two parameters simultaneously, reduced the quercetin extraction yield. The reduction in the contact surface between the solvent and the solid sample and an increase in the concentration of solid load in the sample due to solvent evaporation can lead to a decrease in the extraction yield. In addition, MAE may destroy or change the structure of bioactive compounds due to thermal degradation [47].

The slope of the response surface in the 3-D graph (Fig. 4D) analyses the effects of ethanol concentration and ultrasound temperature on the quercetin extraction yield. Hence, at 80 °C with decreasing ethanol concentration, the response surface showed a decreasing trend for extraction yield. However, at lower temperatures, the yield increased with enhancing the ethanol concentration. In general, reducing the ultrasound temperature and enhancing the amount of solvent at the same time improved the surface of the extraction yield. High temperature can decompose the structure of bioactive compounds and flavonoids in the extracted sample; therefore, it results in a low extraction yield [48].

The response surfaces and contour plots for total flavonoids extraction show the interaction of extraction factors (Fig. 5A-D). The response surface in Fig. 5A indicates the interaction between the ultrasound temperature and the ultrasound time. The rate of flavonoid extraction decreased as the two factors increased simultaneously. In addition, it shows the maximum value of extraction yield for total flavonoids increased at low temperatures, but at temperatures above 70 °C the yield decreased. The explanation for this observation is that high extraction temperatures may degrade the structure of flavonoids during the experiment [49].

In the 3-D graph with consideration of the microwave time (Fig. 5B), it is observed that increasing the microwave time led to an enhancement in the extraction yield when the ultrasound time was at its maximum value of 45 min. But in general, the maximum value of the total flavonoids extraction yield reduced with an increase in the microwave time. This is because the long time of exposure to microwave irradiation and immediately thereafter, being exposed to ultrasonic waves can destroy the sensitive compounds in the sample [50]. Also, reducing the extraction time to the optimum value and achieving a desirable fit between the two extraction times can increase the response surface of the total flavonoids extraction yield.

The response surface which exhibits the interaction between microwave time and ultrasound temperature on the yield of total flavonoids extraction is depicted in Fig. 5C. It indicates that the maximum yield occurs at high microwave time and low ultrasound temperature. But by raising the temperature to 80 ºC, the increase of the microwave time resulted in a decrease in the extraction yield. Despite the fact that increasing the temperature enhances the mass transfer and recovery yield, but the use of high ultrasound temperature can damage the structure of sensitive flavonoids and reduce the extraction yield [51].

The slope of the response surface in Fig. 5D shows that the maximum total flavonoids yield coincides with an enhance in the ethanol concentration and a decrease in the ultrasound temperature. It is also observed that at very high temperatures, as the ethanol concentration increases, the recovery yield also increases. Hence, it indicates a significant interaction of these two parameters on the response surface in the process of total flavonoids extraction.

**Fig. 4**

**Fig. 5**

## 4. Validation of the model and the MUAE efficiency

The maximum extraction yield of quercetin at the predicted conditions (10.85%) was achieved at an ultrasound time of 15.49 min, ultrasound temperature of 67.31 ºC, solvent-to-solid ratio of 20.33 mL/g, microwave time of 46.83 s and ethanol concentration of 69.34%. The predicted condition for the extraction of total flavonoids at ultrasound time of 16.45 min, ultrasound temperature of 64.86 ºC, solvent-to-solid ratio of 28.00 mL/g, microwave time of 38.34 s and ethanol concentration of 69.50% provided the maximum extraction yield of 12.90%. In order to verify the validity of the model, the specific conditions for the maximum responses were predicted using response surfaces with numerical optimization as follows: ultrasound time 15.02 min, ultrasound temperature 61.56 ºC, solvent-to-solid ratio 39.87 mL/g, microwave time 40.57 s and ethanol concentration 62.43%. For consideration of practical limitations and convenient operation, the extraction factors were adjusted to be 15.00 min, 60 ºC, 40 mL/g, 40 s and 60%. The predicted values for recovery yields were 10.05% for quercetin and 12.72% for total flavonoids. These predicted points were very close to the experimental results for $Y_1$ (10.01%) and $Y_2$ (12.36%), with relative percent deviations of 0.39 and 2.83%, respectively. This strong correlation signifies the suitability of the model to predict the responses at optimal conditions. This study suggests these optimal conditions could be significantly useful for further scale-up of the MUAE process for quercetin and total flavonoids extraction from red onion skin.

5. **Comparison of the utilized extraction methods**

Different extraction methods (MAE, UAE, UMAE and MUAE) were used in this study to extract quercetin and total flavonoids from red onion skin. Optimum extraction conditions including microwave time of 60 s, ultrasound time of 15 min, ultrasound temperature of 70 °C, ethanol concentration of 70%, solvent-to-solid ratio of 30 mL/g, microwave power of 180W, ultrasound power of 100%, ultrasound frequency of 37 kHz and particle size of 180 μm were applied for quercetin recovery. In addition, extraction conditions at microwave time of 60 s, ultrasound time of 15 min, ultrasound temperature of 60 ºC, ethanol concentration of 60%, solvent-to-solid ratio of 30 mL/g, microwave power of 180W, ultrasound power of 100%, ultrasound frequency of 37 kHz and particle size of 180 μm were investigated for total flavonoids extraction. The results (Table 4) showed that the highest extraction yield was obtained from the MUAE process, followed by UMAE, UAE and MAE. Also, the classification of different flavonoids content in the extract samples is shown in Table 5.

**Table 4**

**Table 5**

6. **Conclusion**

The purpose of this study was to optimize the extraction conditions in the MUAE process in order to maximize the recovery yields of quercetin and total flavonoids from DOS. According to the results, the optimum quercetin extraction condition for the five effective factors was obtained at microwave time of 60 s, ultrasound time of 15 min, ultrasound temperature of 70 °C, ethanol concentration of 70%, and solvent-to-solid ratio of 30 mL/g. According to the results, ultrasound temperature was found to be an influential factor for the recovery of both compounds. Maximum obtained recovery yield of quercetin and total flavonoids was 10.32 and 12.52%, respectively. The recovery yields obtained from different extraction methods were as follows: UMAE (7.66% quercetin and 10.18% total flavonoids), UAE (5.36% quercetin and 8.34% total flavonoids), and MAE (5.03% quercetin and 7.91 % total flavonoids). The highly significant results of RSM indicate that the model is valid for the extraction of quercetin and total flavonoids and the method is suitable for predicting responses under optimal conditions. This study suggests that the model could be potentially beneficial for further scale-up in the food processes to recover valuable products from red onion skin waste.

**Declarations**

**Funding:** The present research was supported by Biotechnology Research Laboratory, Babol Noshirvani University of Technology, Iran.

**Conflicts of interest/Competing interests:** The authors declare no conflicts of interest.

**Availability of data and material:** Not applicable.

**Code availability:** Not applicable.

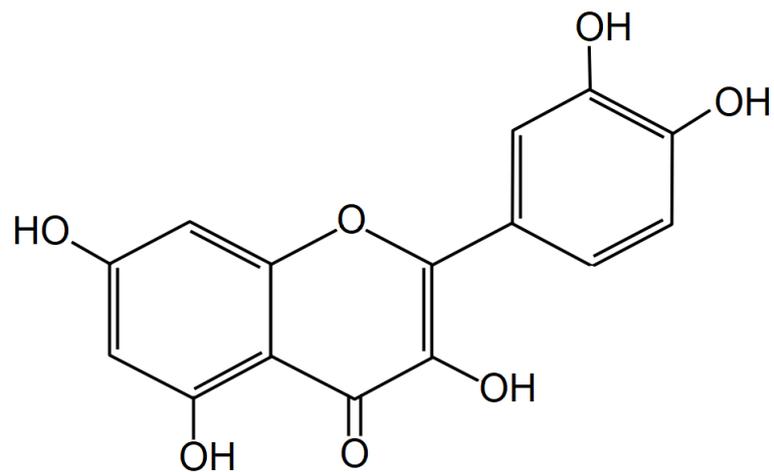

**Fig. 1.** Chemical structure of quercetin

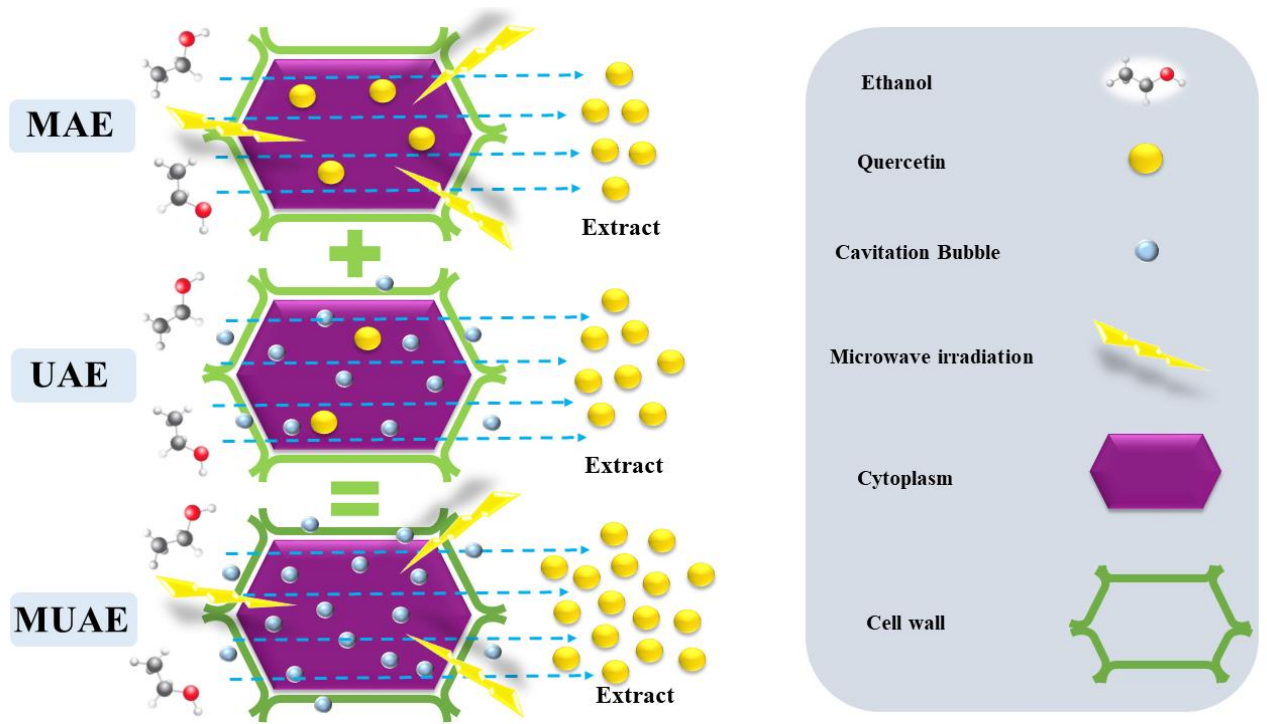

**Fig. 2.** The comparison of the recovery method of the three extraction techniques

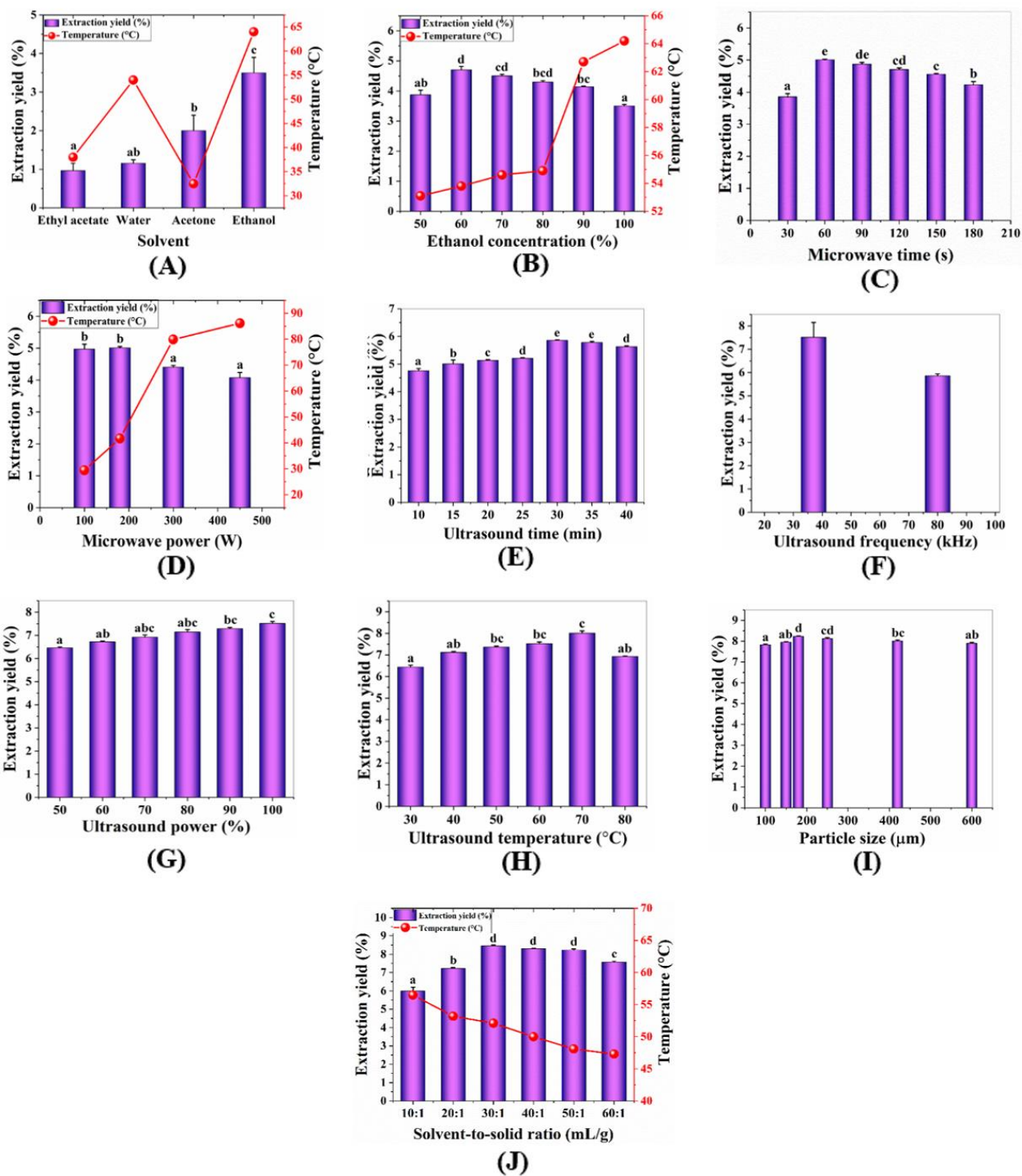

**Fig. 3.** (A–J) Effects of MUAE factors on the yield of quercetin (% = mg QE /g DOS × 100) from red onion skin in single-factor experiments: solvent type (A); ethanol concentration (B); microwave time (C); microwave power (D); ultrasound time (E); ultrasound frequency (F); ultrasound power (G); ultrasound temperature (H); particle size (I) and solvent-to-solid ratio (J). Data are expressed as means ± SD (n=3); mean values with different letters are significantly different ($p < 0.05$)

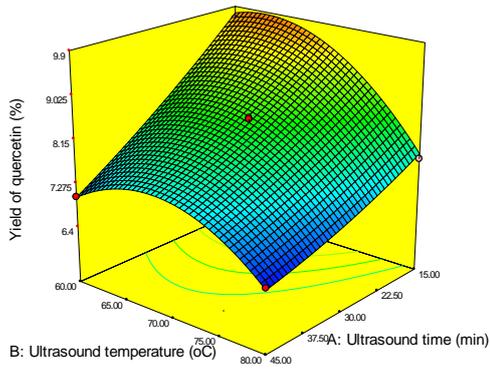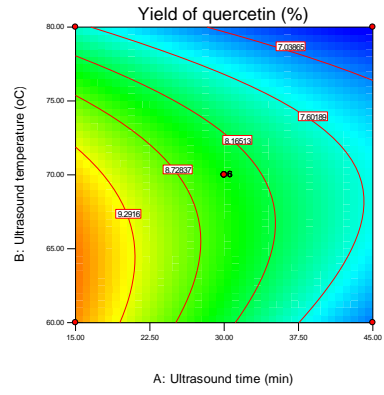

(A)

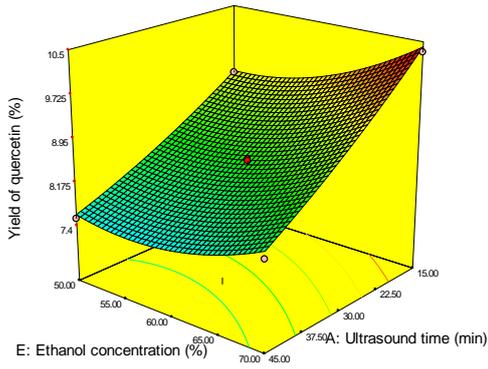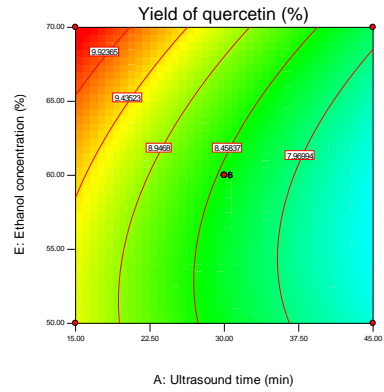

(B)

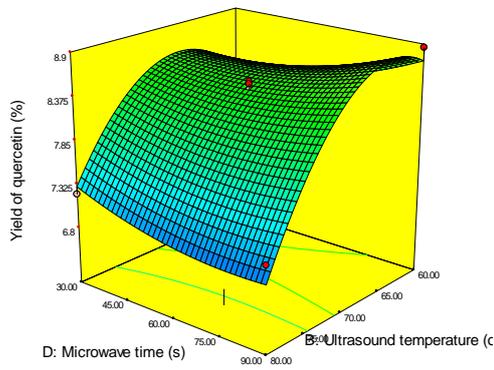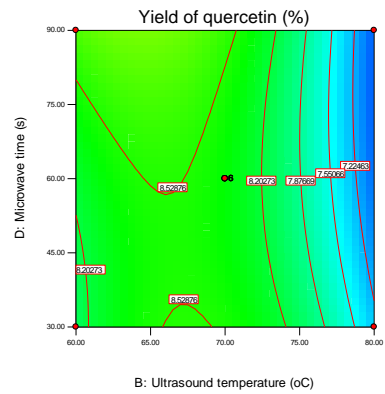

(C)

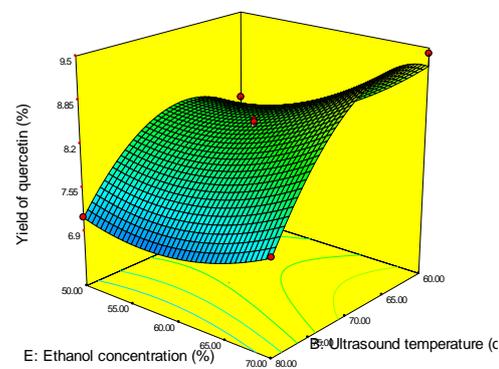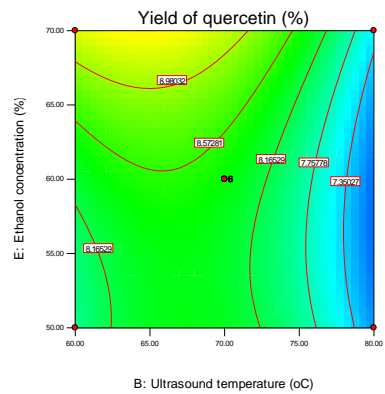

(D)

**Fig. 4.** (A–D). Response surface and contour plots for the MUAE of quercetin from red onion skin wastes with respect to ultrasound temperature and ultrasound time (A); ethanol concentration and ultrasound time (B); microwave time and ultrasound temperature (C) and ethanol concentration and ultrasound temperature (D).

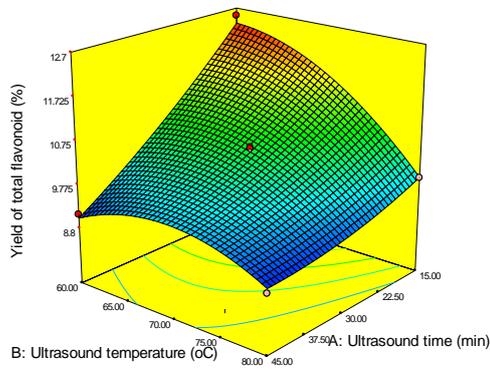
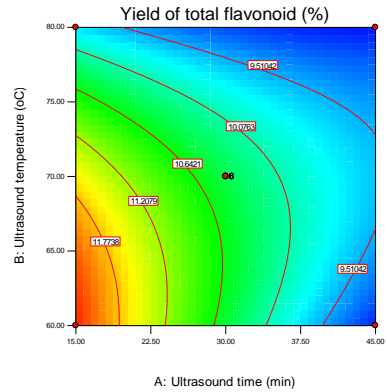

**(A)**

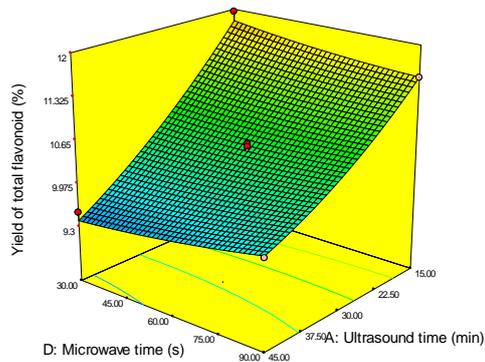
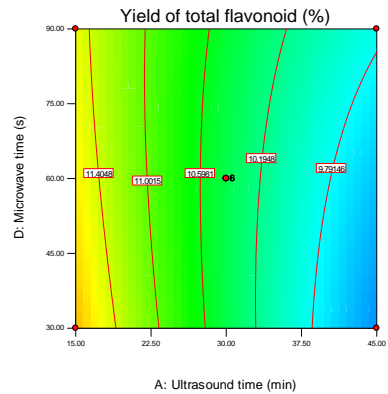

**(B)**

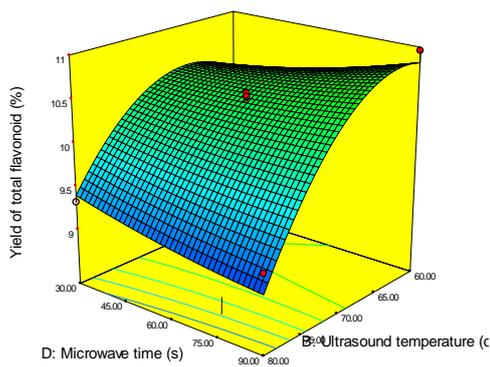
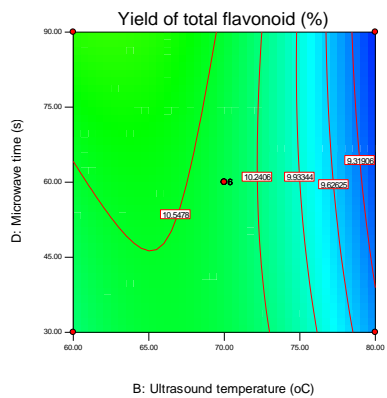

**(C)**

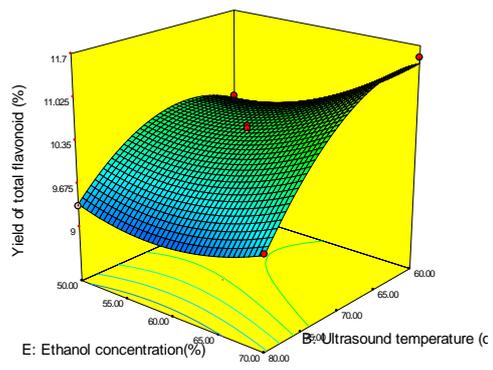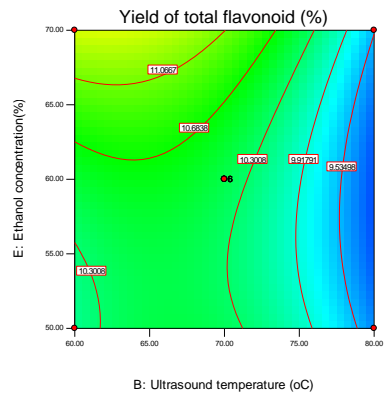

**(D)**

**Fig. 5.** (A–D). Response surface and contour plots for the MUAE of total flavonoids from red onion skin wastes with respect to ultrasound temperature and ultrasound time (A); microwave time and ultrasound time (B); microwave time and ultrasound temperature (C) and ethanol concentration and ultrasound temperature (D).

**Table 1**

Specifications of the independent variables.

| Variables | Codes | Units | Levels | | |
|---|---|---|---|---|---|
| | | | −1 | 0 | +1 |
| Ultrasound time | $X_1$ | min | 15 | 30 | 45 |
| Ultrasound temperature | $X_2$ | °C | 60 | 70 | 80 |
| Solvent-to-solid ratio | $X_3$ | mL/g | 20 | 30 | 40 |
| Microwave time | $X_4$ | s | 30 | 60 | 90 |
| Ethanol concentration | $X_5$ | % | 50 | 60 | 70 |

**Table 2**

Box–Behnken design with coded independent variables and responses.

| Run | Independent variables | | | | | Responses (dependent variables) | |
|---|---|---|---|---|---|---|---|
| | $X_1$(min) | $X_2$(°C) | $X_3$(mL/g) | $X_4$(s) | $X_5$(%) | $Y_1$ (%) | $Y_2$ (%) |
| 1 | -1 | -1 | 0 | 0 | 0 | 9.63 | 12.52 |
| 2 | 0 | -1 | 0 | 0 | 1 | 9.42 | 11.52 |
| 3 | 0 | -1 | -1 | 0 | 0 | 8.41 | 10.41 |
| 4 | 1 | 0 | 0 | -1 | 0 | 7.64 | 9.52 |
| 5 | 0 | 0 | 0 | 0 | 0 | 8.51 | 10.49 |
| 6 | 0 | 0 | 1 | -1 | 0 | 8.79 | 10.74 |
| 7 | 0 | 0 | 0 | 0 | 0 | 8.35 | 10.53 |
| 8 | -1 | 0 | 0 | 0 | -1 | 9.21 | 11.32 |
| 9 | -1 | 0 | 0 | 1 | 0 | 9.43 | 11.50 |
| 10 | 1 | -1 | 0 | 0 | 0 | 7.01 | 9.11 |
| 11 | 0 | 1 | 0 | -1 | 0 | 7.21 | 9.32 |
| 12 | 0 | 0 | 1 | 0 | 1 | 9.51 | 11.74 |
| 13 | 1 | 0 | 0 | 0 | 1 | 7.96 | 9.94 |
| 14 | 0 | 0 | -1 | 1 | 0 | 8.75 | 10.69 |
| 15 | 0 | 0 | -1 | -1 | 0 | 8.62 | 10.53 |
| 16 | 0 | 1 | 1 | 0 | 0 | 7.27 | 9.44 |
| 17 | 1 | 1 | 0 | 0 | 0 | 6.51 | 8.84 |
| 18 | 0 | 0 | 0 | 1 | -1 | 8.42 | 10.30 |
| 19 | 0 | 0 | 0 | 0 | 0 | 8.40 | 10.48 |
| 20 | 0 | 0 | 0 | 1 | 1 | 9.19 | 11.08 |
| 21 | -1 | 1 | 0 | 0 | 0 | 7.58 | 9.71 |
| 22 | -1 | 0 | 0 | -1 | 0 | 9.92 | 11.96 |
| 23 | 0 | 1 | 0 | 1 | 0 | 7.13 | 9.24 |
| 24 | 1 | 0 | 0 | 0 | -1 | 7.52 | 9.78 |
| 25 | 0 | 0 | 1 | 1 | 0 | 9.16 | 10.75 |
| 26 | 0 | 0 | -1 | 0 | 1 | 9.46 | 11.35 |
| 27 | 0 | 1 | -1 | 0 | 0 | 7.23 | 9.40 |
| 28 | -1 | 0 | 0 | 0 | 1 | 10.32 | 12.03 |
| 29 | 1 | 0 | 1 | 0 | 0 | 7.95 | 9.93 |
| 30 | 0 | 0 | 1 | 0 | -1 | 8.61 | 10.62 |
| 31 | 0 | 0 | 0 | 0 | 0 | 8.41 | 10.35 |
| 32 | 0 | -1 | 0 | 1 | 0 | 8.86 | 10.96 |
| 33 | -1 | 0 | 1 | 0 | 0 | 10.01 | 12.01 |
| 34 | 0 | -1 | 0 | -1 | 0 | 7.95 | 10.11 |
| 35 | 0 | 0 | 0 | -1 | 1 | 9.34 | 11.16 |
| 36 | -1 | 0 | -1 | 0 | 0 | 9.98 | 11.91 |
| 37 | 0 | -1 | 0 | 0 | -1 | 8.11 | 10.23 |
| 38 | 1 | 0 | 0 | 1 | 0 | 7.77 | 9.81 |
| 39 | 0 | 1 | 0 | 0 | -1 | 7.11 | 9.33 |
| 40 | 1 | 0 | -1 | 0 | 0 | 7.84 | 9.67 |
| 41 | 0 | -1 | 1 | 0 | 0 | 8.52 | 10.79 |
| 42 | 0 | 1 | 0 | 0 | 1 | 7.47 | 9.57 |
| 43 | 0 | 0 | 0 | 0 | 0 | 8.48 | 10.40 |
| 44 | 0 | 0 | 0 | 0 | 0 | 8.31 | 10.25 |
| 45 | 0 | 0 | -1 | 0 | -1 | 8.62 | 10.73 |
| 46 | 0 | 0 | 0 | -1 | -1 | 8.43 | 10.38 |

% = mg QE /g DOS × 100; $X_1$: ultrasound time; $X_2$: ultrasound temperature; $X_3$: solvent-to-solid ratio; $X_4$: microwave time; $X_5$: ethanol concentration; $Y_1$: yield of quercetin; $Y_2$: yield of total flavonoids

**Table 3**

ANOVAs for the regression models.

| Source | | Sum of squares | df. | Mean square | F-value | P-value | Significance |
|---|---|---|---|---|---|---|---|
| Model | $Y_1$ | 36.97 | 20 | 1.85 | 85.15 | < 0.0001 | ** |
| | $Y_2$ | 34.93 | 20 | 1.75 | 65.18 | < 0.0001 | ** |
| X1 | $Y_1$ | 15.76 | 1 | 15.76 | 726.04 | < 0.0001 | ** |
| | $Y_2$ | 16.73 | 1 | 16.73 | 624.25 | < 0.0001 | ** |
| X2 | $Y_1$ | 6.76 | 1 | 6.76 | 311.41 | < 0.0001 | ** |
| | $Y_2$ | 7.29 | 1 | 7.29 | 272.05 | < 0.0001 | ** |
| X5 | $Y_1$ | 2.76 | 1 | 2.76 | 126.94 | < 0.0001 | ** |
| | $Y_2$ | 2.03 | 1 | 2.03 | 75.78 | < 0.0001 | ** |
| X1X2 | $Y_1$ | 0.60 | 1 | 0.60 | 27.67 | < 0.0001 | ** |
| | $Y_2$ | 1.61 | 1 | 1.61 | 60.19 | < 0.0001 | ** |
| $X_1X_4$ | $Y_1$ | 0.096 | 1 | 0.096 | 4.43 | 0.0456 | * |
| | $Y_2$ | 0.14 | 1 | 0.14 | 5.25 | 0.0307 | * |
| $X_1X_5$ | $Y_1$ | 0.11 | 1 | 0.11 | 5.17 | 0.0318 | * |
| | $Y_2$ | 0.076 | 1 | 0.076 | 2.82 | 0.1054 | |
| $X_2X_4$ | $Y_1$ | 0.25 | 1 | 0.25 | 11.29 | 0.0025 | * |
| | $Y_2$ | 0.22 | 1 | 0.22 | 8.07 | 0.0088 | * |
| $X_2X_5$ | $Y_1$ | 0.23 | 1 | 0.23 | 10.39 | 0.0035 | * |
| | $Y_2$ | 0.28 | 1 | 0.28 | 10.29 | 0.0037 | * |
| $X_1^2$ | $Y_1$ | 0.11 | 1 | 0.11 | 5.11 | 0.0328 | * |
| | $Y_2$ | 0.26 | 1 | 0.26 | 9.71 | 0.0046 | * |
| $X_2^2$ | $Y_1$ | 5.48 | 1 | 5.48 | 252.36 | < 0.0001 | ** |
| | $Y_2$ | 2.96 | 1 | 2.96 | 110.43 | < 0.0001 | ** |
| $X_3^2$ | $Y_1$ | 0.88 | 1 | 0.88 | 40.37 | < 0.0001 | ** |
| | $Y_2$ | 0.63 | 1 | 0.63 | 23.49 | < 0.0001 | ** |
| $X_4^2$ | $Y_1$ | 0.17 | 1 | 0.17 | 7.90 | 0.0095 | * |
| | $Y_2$ | 0.019 | 1 | 0.019 | 0.72 | 0.4056 | |
| $X_5^2$ | $Y_1$ | 0.86 | 1 | 0.86 | 39.73 | < 0.0001 | ** |
| | $Y_2$ | 0.78 | 1 | 0.78 | 29.19 | < 0.0001 | ** |
| residual | $Y_1$ | 0.54 | 25 | 0.022 | | | |
| | $Y_2$ | 0.67 | 25 | 0.027 | | | |
| Lack of Fit | $Y_1$ | 0.51 | 20 | 0.026 | 4.49 | 0.0515 | Not significant |
| | $Y_2$ | 0.62 | 20 | 0.031 | 2.81 | 0.1276 | Not significant |

|  | | | | |
|---|---|---|---|---|
| Pure Error | $Y_1$ | 0.029 | 5 | 5.720E-003 |
|  | $Y_2$ | 0.055 | 5 | 1.10E-02 |
| Cor Total | $Y_1$ | 37.51 | 45 | |
|  | $Y_2$ | 35.60 | 45 | |
| Std. Dev. | $Y_1$ | 0.15 | | |
|  | $Y_2$ | 0.16 | | |
| $R^2$ | $Y_1$ | 0.9855 | | |
|  | $Y_2$ | 0.9812 | | |
| Adjusted $R^2$ | $Y_1$ | 0.9740 | | |
|  | $Y_2$ | 0.9661 | | |
| Predicted $R^2$ | $Y_1$ | 0.9441 | | |
|  | $Y_2$ | 0.9287 | | |
| Mean | $Y_1$ | 8.44 | | |
|  | $Y_2$ | 10.49 | | |
| C.V. % | $Y_1$ | 1.75 | | |
|  | $Y_2$ | 1.56 | | |

df: degree of freedom; * significant; ** highly significant.

**Table 4**

Comparison of recovery yields from different extraction methods

| Extraction method | QE (%) | TFC (%) |
|---|---|---|
| MAE | 5.03% ± 0.02[a] | 7.91 % ± 0.06[a] |
| UAE | 5.36% ± 0.0[b] | 8.34% ± 0.05[b] |
| UMAE | 7.66% ± 0.04[c] | 10.18% ± 0.14[c] |
| MUAE | 10.32% ± 0.18[d] | 12.52% ± 0.20[d] |

% = mg QE /g DOS × 100; MAE: microwave assisted extraction; UAE: ultrasound assisted extraction; UMAE: ultrasound-microwave assisted extraction; MUAE: microwave-ultrasound assisted extraction; TFC: total flavonoids content; QE: quercetin equivalents; data are expressed as means ± SD (n=3); mean values with different letters are significantly different ($p < 0.05$).

**Table 5**

Flavonoids content in extracts of red onion skin by HPLC

| Compound | Chemical formula | Molecular weight (g/mol) | Retention time (min) | Concentration (%) |
|---|---|---|---|---|
| Quercetin 7,4'-O-diglucoside | $C_{27}H_{30}O_{17}$ | 626.5 | 7.10 | $0.65 \pm 0.02^a$ |
| Quercetin 3,4'-O-diglucoside | $C_{27}H_{30}O_{17}$ | 626.5 | 7.40 | $0.73 \pm 0.02^a$ |
| Isorhamnetin 3,4'-diglucoside | $C_{28}H_{32}O_{17}$ | 640.5 | 7.60 | ND |
| Unknown | | | 8.20 | ND |
| Quercetin 3-O-glucoside | $C_{21}H_{20}O_{12}$ | 464.4 | 8.44 | $0.36 \pm 0.01^a$ |
| Quercetin-4'-O-glucoside | $C_{21}H_{20}O_{12}$ | 464.4 | 10.50 | $3.15 \pm 1.20^b$ |
| Isorhamnetin-3-O-glucoside | $C_{22}H_{22}O_{12}$ | 478.4 | 12.60 | ND |
| Quercetin | $C_{15}H_{10}O_7$ | 302.23 | 13.60 | $5.43 \pm 0.06^c$ |
| kaempferol | $C_{15}H_{10}O_6$ | 286.24 | 16.50 | $0.46 \pm 0.08^a$ |
| Isorhamnetin | $C_{16}H_{12}O_7$ | 316.26 | 16.80 | ND |
| Unknown | | | 17.20 | ND |

% = mg QE /g DOS × 100; ND: not detected; data are expressed as means ± SD (n=3); mean values with different letters are significantly different ($p < 0.05$).